\documentclass[aps,preprint]{revtex4}%
\usepackage{amsfonts}
\usepackage{amsmath}
\usepackage{amssymb}
\usepackage{graphicx}%
\providecommand{\U}[1]{\protect\rule{.1in}{.1in}}

\begin{document}
\preprint{ }
\title[ ]{Klein-Gordon Equation with Coulomb Potential in the Presence of a Minimal
Length }
\author{Djamil Bouaziz}
\email{djamilbouaziz@univ-jijel.dz}
\affiliation{Laboratoire de Physique Th\'{e}orique, Universit\'{e} de Jijel, BP 98, Ouled
Aissa, 18000 Jijel, Algeria.}

\begin{abstract}
We study the Klein-Gordon equation for Coulomb potential $V(r)=(-Ze%
{{}^2}%
)/r$ in quantum mechanics with a minimal length. The zero energy solution is
obtained analytically in momentum space in terms of Heun's functions. The
asymptotic behavior of the solution shows that the presence of a minimal
length regularize the potential in the strong attractive regime, $Z>68$. The
equation with nonzero energy is established in a particular case in the first
order of the deformation parameter; it is a generalized Heun's equation.

\end{abstract}
\volumeyear{2013}
\volumenumber{number}
\issuenumber{number}
\eid{identifier}
\date[Date text]{date}
\received[Received text]{date}

\revised[Revised text]{date}

\accepted[Accepted text]{date}

\published[Published text]{date}

\startpage{1}
\maketitle

\section{Introduction}

In recent years, a lot of attention has been attracted to the study of
physical problems within the formalism of quantum mechanics with a generalized
uncertainty relation \cite{1}, including a minimal length, see, for instance
Refs. \cite{2}, and for a recent detailed review and a large list of
references in connection with this subject, see, Ref. \cite{hossenfelder}.
This modified version of quantum mechanics is based on the following deformed
commutation relations between position and momentum operators \cite{k,3}:%
\begin{align}
\lbrack\widehat{X}_{i},\widehat{P}_{j}] &  =i\hbar\lbrack(1+\beta\widehat
{P}^{2})\delta_{ij}+\beta^{^{\prime}}\widehat{P}_{i}\widehat{P}_{j}],\text{
\ \ \ \ \ \ \ }(\beta,\beta^{^{\prime}})>0,\nonumber\\
\lbrack\widehat{P}_{i},\widehat{P}_{j}] &  =0,\text{ \ \ }[\widehat{X}%
_{i},\widehat{X}_{j}]=i\hbar\frac{2\beta-\beta^{^{\prime}}+\beta(2\beta
+\beta^{^{\prime}})\widehat{P}^{2}}{1+\beta\widehat{P}^{2}}\left(  \widehat
{P}_{i}\widehat{X}_{j}-\widehat{P}_{j}\widehat{X}_{i}\right)  \text{.}%
\label{1}%
\end{align}

These commutators lead to the, so-called, generalized uncertainty principle
(GUP), which implies the existence of a nonzero minimal uncertainty in
position (minimal length) given, in $N$ dimensions, by \cite{k}%
\begin{equation}
\left(  \Delta X_{i}\right)  _{\min}=\hbar\sqrt{\left(  N\beta+\beta
^{^{\prime}}\right)  },\text{ \ \ }\forall i.
\end{equation}

The idea of modifying the Heiseinberg uncertainty relation in such a way that
it incorporate a minimal length has been, first, proposed in quantum gravity
and string theory \cite{garay,amati,m}, where the minimal length is supposed
to be on the order of the Planck scale. However, it was argued that in
nonrelativistic or relativistic quantum mechanics \cite{k,4}, the minimal
length may be viewed as an intrinsic scale characterizing the system under
study. Furthermore, it was shown in Ref. \cite{4} that this elementary length
regularize in a natural way the strong attractive inverse square potential,
known to be singular in quantum mechanics.

In this work, we study the Klein-Gordon (KG) equation for Coulomb potential,
$V(r)=(-Ze%
{{}^2}%
)/r$, in the presence of a minimal length. In ordinary KG equation, this
problem becomes singular when $Z>68$, and the potential must be regularized by
introducing a cutoff and modifying the interaction at short distances
\cite{5}. It is thus interesting to examine to what extend the introduction of
a minimal length in the formalism regularizes the strong attractive Coulomb potential.

\section{Ordinary KG equation for Coulomb potential: Momentum space treatment}

In QM with a minimal length, momentum space is more convenient that coordinate
space \cite{1}. So, for the sake of further discussion, we give here the
solution to ordinary KG equation for Coulomb potential in the momentum
representation. We shall be interested in the singularity structure of this
equation to illustrate how the strong attractive potential becomes singular.
The KG equation for the potential $V(r)=(-Ze%
{{}^2}%
)/r$\ reads%
\begin{equation}
\left(  E\widehat{R}^{2}+Ze^{2}E\widehat{R}+Ze^{2}\right)  \psi\left(
p\right)  =\widehat{R}^{2}\left(  m^{2}c^{4}+p^{2}c^{2}\right)  \psi\left(
p\right)  . \label{2}%
\end{equation}

In the case of zero angular momentum quantum number ($\ell=0$), the distance
squared and distance operators act in momentum space as%
\[
R^{2}\psi\left(  p\right)  =-\hbar^{2}\left(  \frac{d^{2}}{dp^{2}}+\frac{2}%
{p}\frac{d}{dp}\right)  \psi\left(  p\right)  ,\text{ \ \ \ \ }R\psi\left(
p\right)  =i\hbar\left(  \frac{d}{dp}+\frac{1}{p}\right)  \psi\left(
p\right)  .
\]

Replacing in Eq. (\ref{2}), we get%
\begin{align}
&  \hbar^{2}\left(  \epsilon^{2}+c^{2}p^{2}\right)  \frac{d^{2}\psi\left(
p\right)  }{dp^{2}}+\left(  \frac{2}{p}\hbar^{2}\epsilon+2i\hbar
ZEe^{2}+6\hbar^{2}c^{2}p\right)  \frac{d\psi\left(  p\right)  }{dp}%
+\nonumber\\
&  +\left(  Z^{2}e^{4}+\frac{2i\hbar ZEe^{2}}{p}+6\hbar^{2}c^{2}\right)
\psi\left(  p\right)  =0. \label{3}%
\end{align}
where $\epsilon^{2}=m^{2}c^{4}-E^{2}.$

The two solutions of this equation behave at infinity as%
\begin{equation}
\psi_{1}\left(  p\right)  \backsim p^{-\frac{5}{2}-\mu}\text{,
\ \ \ \ \ \ \ \ }\psi_{2}\left(  p\right)  \backsim p^{-\frac{5}{2}+\mu},
\label{4}%
\end{equation}
where $\mu=\sqrt{\frac{1}{4}-\frac{e^{4}Z^{2}}{\hbar^{2}c^{2}}}.$

If Z $<68$, the first solution falls off more rapidly than the other, and thus
it is selected as the physical solution. It can also be shown that the average
value of the operator $\widehat{R}^{2}$ is divergent at infinity
($\left\langle \widehat{R}\right\rangle _{\psi_{2}}\rightarrow\infty$)
\cite{bound state}. When $Z<68$, the general solution is a linear combination
of the two solutions that behave in the same manner. Accordingly, the wave
function will depend on an arbitrary phase parameter, which is a common
feature of singular potentials \cite{case}.

The complete solution to Eq. (\ref{3}) is written in terms of the
hypergeometric function as%
\begin{equation}
\psi\left(  p\right)  =Ap^{-1}\left(  1+(ic/\epsilon)p\right)  ^{-\frac{3}%
{2}-\mu}F\left(  \frac{3}{2}+\mu,\frac{1}{2}-w+\mu,2\mu+1;2/(1+(ic/\epsilon
)p)\right)  ,\text{\ } \label{5}%
\end{equation}
where $F$ is a hypergeometric function, $A$ is a normalization constant and
$w=\frac{ZEe^{2}}{\hbar c\epsilon}.$

The energy spectrum can be obtained by setting%
\begin{equation}
1/2-w+\mu=-n,\text{ \ \ \ \ \ \ \ \ }n=0,1,2..., \label{6}%
\end{equation}
to ensure that the wave function (\ref{5}) be square integrable. In this case,
the hypergeometric series reduces to a polynomial.

Equation (\ref{6}) constitutes the quantization condition of the energy, which
give the well-known discrete energy spectrum of Coulomb potential \cite{5}.

In the case $Z>68$, the parameter $%
\mu
$ becomes imaginary and thus the spectral condition (\ref{6}) fails. In order
to obtain a discrete spectrum, the potential must be regularized by
introducing a cutoff at short distances \cite{5}. Another alternative approach
was discussed in Ref. \cite{alha} to deal with the strong attractive Couomb
potential in the context of Dirac equation to overcome the dependence of the
problem on the numbre\ $Z.$

In the following section we will show that, when a minimal length is
introduced in the K-G equation, there will be any difference between the
strong and weak attractive regimes.

\section{KG equation for Coulomb potential with a minimal length}

In the literature, one of the most used representations of the position and
momentum operators satisfying Eqs. (\ref{1}) is \cite{3,4}%
\begin{equation}
\widehat{X}_{i}=i\hbar\left[  \left(  1+\beta p^{2}\right)  \frac{\partial
}{\partial p_{i}}+\beta^{\prime}p_{i}p_{j}\frac{\partial}{\partial p_{j}%
}+\gamma p_{i}\right]  ,\text{ \ \ \ \ \ }\widehat{P}_{i}=p_{i}, \label{7}%
\end{equation}
where $\gamma$ is a small positive parameter related to $\beta$,
$\beta^{\prime}$; it does not affect the observable quantities.

In the case\ $\ell=0$ and $\gamma=0$, representation (\ref{7}) leads to the
following expression of the distance squared operator \cite{3,4}:%
\begin{equation}
\widehat{R}^{2}=%
{\textstyle\sum\limits_{i=1}^{3}}
\widehat{X}_{i}\widehat{X}_{i}\mathbf{=}\left(  i\hbar\right)  ^{2}\left\{
\left[  1+\left(  \beta+\beta^{\prime}\right)  p^{2}\right]  ^{2}\frac{d^{2}%
}{dp^{2}}+\frac{2}{p}\left[  1+\left(  \beta+\beta^{\prime}\right)
p^{2}\right]  \left[  1+\left(  2\beta+\beta^{\prime}\right)  p^{2}\right]
\frac{d}{dp}\right\}  . \label{8}%
\end{equation}

As it was shown in Sec. 2, the singularity of the strong attractive Coulomb
potential manifests at infinity in momentum space (or equivalently, at short
distances in coordinate space). In this limit, the solution of the K-G
equation does not depend on the energy $E$. It is thus interesting to begin
our study by examining the effect of the minimal length on the zero-energy K-G equation.

\subsection{The zero energy case}

Inserting Eq. (\ref{8}) in the KG equation (\ref{2}), we obtain after some
calculation the following equation:
\[
\frac{d^{2}\psi\left(  p\right)  }{dp^{2}}+\frac{2}{p}\left\{  4\left[
\frac{p^{2}+\frac{1}{2}m^{2}c^{2}}{p^{2}+m^{2}c^{2}}\right]  -\frac
{1+\beta^{\prime}p^{2}}{1+\left(  \beta+\beta^{\prime}\right)  p^{2}}%
\frac{d\psi\left(  p\right)  }{dp}\right\}
\]%
\begin{equation}
+\left\{  \frac{6+\left(  10\beta+6\beta^{\prime}\right)  p^{2}}{\left[
1+\left(  \beta+\beta^{\prime}\right)  p^{2}\right]  }+\frac{Z^{2}e^{4}%
/\hbar^{2}c^{2}}{\left[  1+\left(  \beta+\beta^{\prime}\right)  p^{2}\right]
^{2}}\right\}  \frac{\psi\left(  p\right)  }{p^{2}+m^{2}c^{2}}\text{ }=0.
\label{9}%
\end{equation}
At infinity, the two solutions of this equation behave as
\begin{equation}
\psi_{1}\left(  p\right)  _{p\rightarrow\infty}\backsim p^{-3-2\beta
/(\beta+\beta^{\prime})},\text{ \ \ }\psi_{2}\left(  p\right)  _{p\rightarrow
\infty}\backsim p^{-2}. \label{10}%
\end{equation}
These behaviors are completely different from that of the non deformed case,
see Eq. (\ref{4}). Both solutions are independent of the atomic number $Z$;
moreover, the second solution $\psi_{2}$ does not depend on the deformation
parameters and falls off more slowly than $\psi_{1}$. It follows that
$\psi_{1}$ is manifestly the physical solution regardless the value of $Z$.
Consequently, in the presence of a minimal length, there is no difference
between the strong ($Z>68$) and the weak ($Z<68$) regimes of the potential.
This might be interpreted as a signal of regularization of the potential.

Let us now end this section by mentioning that Eq. (\ref{9}) can be
transformed to the following Heun's differential equation, by using the change
of variable $\xi=\frac{(\beta+\beta^{\prime})p^{2}}{1+(\beta+\beta^{\prime
})p^{2}}$, and the transformation $\psi\left(  \xi\right)  =(1-\xi)f\left(
\xi\right)  $:%
\begin{equation}
\frac{d^{2}f\left(  \xi\right)  }{d\xi^{2}}+\left(  \frac{c}{\xi}+\frac{e}%
{\xi-1}+\frac{d}{\xi-\xi_{0}}\right)  \frac{df\left(  \xi\right)  }{d\xi
}+\left(  \frac{ab\xi+q}{\xi\left(  \xi-1\right)  \left(  \xi-\xi_{0}\right)
}\right)  f\left(  \xi\right)  =0,\label{11}%
\end{equation}
with the parameters%
\begin{align}
a &  =\frac{1}{2}\left(  3-\omega_{1}-\nu\right)  ,\text{ }b=\frac{1}%
{2}\left(  3-\omega_{1}+\nu\right)  ,\text{ }c=\frac{3}{2},\text{ }d=2,\text{
}e=\frac{1}{2}-\omega_{1},\text{ }\nonumber\\
\text{ \ \ \ \ }\omega_{1} &  =\frac{2\beta}{\beta+\beta^{\prime}}%
,\text{\ }\nu=\left[  \left(  \omega_{1}-1\right)  ^{2}-\frac{4k}%
{1-2\omega_{2}}\right]  ^{\frac{1}{2}},\text{ }q=-\left(  \frac{3}{2}+\frac
{k}{1-2\omega_{2}}\right)  ,\text{ }\nonumber\\
\xi_{0} &  =\frac{2\omega_{2}}{2\omega_{2}-1},\text{ \ \ \ \ \ \ \ \ \ }%
k=\frac{Z^{2}e^{4}}{4\hbar^{2}c^{2}},\text{ \ \ \ \ \ }\omega_{2}=\frac{1}%
{2}\left(  \beta+\beta^{\prime}\right)  m^{2}c^{2}%
\end{align}
The solution to Eq. (\ref{11}), which is regular at $\xi=0$ is \cite{ronv}:%

\begin{equation}
\psi\left(  \xi\right)  =A\left(  1-\xi\right)  H\left(  \xi_{0}%
,q,a,b,c,d,\xi\right)  . \label{12}%
\end{equation}

It is important to mention that solution (\ref{12}) reduces to an
hypergeometric function in the particular case $\beta=\beta^{\prime}$ as
follow:
\begin{equation}
\psi_{\beta=\beta^{^{\prime}}}\left(  \xi\right)  =A\left(  1-\xi\right)
F\left(  a,b,c;\xi/\xi_{0}\right)  , \label{sol}%
\end{equation}
with the parameters
\[
a=5/4-\nu/2,\text{ }b=5/4+\nu/2,\text{ }c=3/2,\text{ }\nu=\sqrt
{1/4-4k/(1-2\omega_{2})}.
\]

\subsection{The nonzero energy case}

The KG equation (\ref{2}) can not be established in the presence of a minimal
length, because the definition of the operator $\widehat{R}$ is not obvious as
long as the $\widehat{R}^{2}$ operator, given by Eq. (\ref{8}), is not
factorizable in general. However, it was shown in Ref. \cite{6} that, in the
particular case $\beta^{\prime}$ $=2\beta$, this operator can be factorized in
the first order of the deformation parameter, and thus the KG equation can be
obtained in this special case. In fact, by using \cite{6}%
\begin{align}
\widehat{R}^{2} &  \mathbf{=}\left(  i\hbar\right)  ^{2}\left\{  (1+6\beta
p^{2})\frac{d^{2}}{dp^{2}}+\frac{2}{p}\allowbreak(1+7\beta p^{2})\frac{d}%
{dp}\right\}  +O\left(  \beta^{2}\right)  ,\\
\widehat{R} &  \mathbf{=}i\hbar\left[  \left(  1+3\beta p^{2}\right)  \frac
{d}{dp}+\frac{1}{p}\allowbreak\allowbreak\left(  1+\beta p^{2}\right)
\right]  +O\left(  \beta^{2}\right)  ,
\end{align}
equation (\ref{2}) takes the form%
\[
(p^{2}+\epsilon^{2})(1+6\beta p^{2})\dfrac{d^{2}\varphi(p)}{dp^{2}}+\left\{
2\beta p(p^{2}+\epsilon^{2})+4p(1+6\beta p^{2})+2i\omega(1+3\beta
p^{2})\right\}  \dfrac{d\varphi(p)}{dp}%
\]%
\begin{equation}
\left\{  -2\beta(p^{2}+\epsilon^{2})-2(1+6\beta p^{2})+4(1+7\beta
p^{2})-4i\omega\beta p+k\right\}  \varphi(p)=0,\label{13}%
\end{equation}
where
\[
\psi(p)=\frac{1}{p}\varphi(p),\text{ \ \ }\omega=\frac{Ze^{2}E}{\hbar c^{2}%
},\text{ \ \ }\epsilon^{2}=m^{2}c^{2}-\frac{E^{2}}{c^{2}},\text{
\ \ \ }k=(\frac{Ze^{2}}{\hbar c})^{2}\ .
\]

By making the change of variable,
\[
x=\frac{1}{2}\left(  1-i\sqrt{6\beta}p\right)  ,
\]
equation (\ref{13}) can be transformed to the following generalized Heun's
equation \cite{hon}:%
\begin{equation}
\dfrac{d^{2}\varphi}{dx^{2}}+\left(  \dfrac{c}{x}+\dfrac{d}{(x-1)}+\dfrac
{e}{(x-x_{1})}+\dfrac{f}{(x-x_{2})}\right)  \dfrac{d\varphi}{dx}+\left(
\dfrac{abx^{2}+\rho_{1}x+\rho_{2}}{x\left(  x-1\right)  (x-x_{1})(x-x_{2}%
)}\right)  \varphi=0,\label{14}%
\end{equation}
With the Fuchsian condition,
\[
a+b+1=c+d+e+f.
\]
The parameters of Eq. (\ref{14}) are given by%
\begin{align*}
a &  =1,\text{ }b=\frac{7}{3},\text{ }\rho_{1}=\left(  -\frac{7}{3}%
-\frac{\omega\sqrt{6\beta}}{3}\right)  ,\text{ }\rho_{2}=\frac{\beta
\epsilon^{2}}{2}+\frac{\omega\sqrt{6\beta}}{6}+\allowbreak\frac{1}{12}%
-\frac{k}{4},\\
c &  =\left(  \frac{1}{6}+\frac{\omega\sqrt{6\beta}}{2\left(  1-6\beta
\epsilon^{2}\right)  }\right)  ,\text{ }d=\frac{1}{6}-\frac{\omega\sqrt
{6\beta}}{2\left(  1-6\beta\epsilon^{2}\right)  },\text{ }e=\left(
2+\frac{\omega\left(  1-3\beta\epsilon^{2}\right)  }{\left(  1-6\beta
\epsilon^{2}\right)  \epsilon}\right)  ,\\
f &  =\left(  2-\frac{\omega\left(  1-3\beta\epsilon^{2}\right)  }{\left(
1-6\beta\epsilon^{2}\right)  \epsilon}\right)  ,\text{ \ \ \ \ \ \  }%
x_{1}=\left(  \frac{1}{2}+\frac{\epsilon\sqrt{6\beta}}{2}\right)  ,\text{
\ \ \ \  }x_{2}=\frac{1}{2}-\frac{\epsilon\sqrt{6\beta}}{2}.
\end{align*}
Equation (\ref{14}) belongs to the class of Fuchsian equations: it is a linear
homogeneous second-order differential equation with five singularities $z=0,$
$1,$ $x_{1},$ $x_{2},$ $\infty$, all regular. So, it admits power series
solutions in the neighborhood of each singular point. However, to the best of
our knowledge, the analytic solutions to the generalized Heun's equation
(\ref{14}) are not known in the literature, i.e., the recurrence relation that
determines the coefficients of the series was not established for equations of
type (\ref{14}). It follows that the formulation of a physical problem with
this kind of equation is interesting in it self. This might motivate profound
studies on such Fuchsian equations with five singularities.

To end this section, let us discuss the effect of the minimal length in the
singularity structure of Eq. (\ref{14}). It can be cheeked that, in the limit
$p\gg1$, the two linearly independent solutions are
\[
\psi_{1}\sim p^{-10/3},\text{ \ \ \ \ \ \ \ \ \ }\psi_{2}\sim\ p^{-2}.
\]
\ This result confirms what we have obtained in the zero energy case; the two
solutions do not have the same behavior as in the ordinary case. Consequently
we can always reject $\psi_{2}$ even for the strong attractive regime $Z>68$,
so that the wave function will not contain an arbitrary phase, characterizing
singular potentials \cite{case}. Remains to mention that the difference
between the physical solution $\psi_{1}$ and the zero energy solution
(\ref{4}) is due to the fact that Eq. (\ref{13}) is obtained in the first
order in $\beta$, however the zero energy equation (\ref{9}) is exact.

\section{ Summary}

We have studied the KG equation for Coulomb potential in quantum mechanics
with a minimal length. The zero energy solution is obtained analytically in
momentum space; it is a Heun's function, which reduces to a hypergeometric
function in the case $\beta=\beta^{\prime}$. The asymptotic behavior of the
solution at large momenta showed that the presence of a minimal length
regularize the potential in the strong attractive regime. The nonzero energy
equation was established in the particular case $\beta^{\prime}=2\beta$ in the
first order of the deformation parameter. It is transformed to a canonical
form of Fuchsian equations, namely, a generalized Heun equation. The behavior
of the solution confirms the regularizing effect of the minimal length.

\begin{acknowledgments}
This work is supported by the Algerian Ministry of Higher Education and
Scientific Research under the PNR Project No. 8/u18/4327 and the CNEPRU
Project No. D017201600026.
\end{acknowledgments}


\begin{thebibliography}{99}                                                                                               %


\bibitem {1}A. Kempf, et al, Phys. Rev. D 52, 1108 (1995).

\bibitem {2}F. Brau, J. Phys. A 32, 7691 (1999); U. Harbach and S.
Hossenfelder, Phys. Lett. B \textbf{632}, 379 (2006); F. Brau and F.
Buisseret, Phys. Rev. D \textbf{74}, 036002 (2006); D. Bouaziz and M. Bawin,
Phys. Rev. A\textbf{ 78}, 032110 (2008); C. Quesne and V. M. Tkachuk, Phys.
Rev. A \textbf{81}, 012106 (2010); A. Bina, S. Jalalzadeh and A. Moslehi,
Phys. Rev. D \textbf{81}, 023528 (2010); P. Pedram, Physics Letters B 710 478
(2012); T.L. Antonacci Oakes, R.O. Francisco, J.C. Fabris, J.A. Nogueiraa,
Eur. Phys. J. C \textbf{73}, 2495 (2013).

\bibitem {hossenfelder}S. Hossenfelder, Living Rev. Relativity. \textbf{16}, 2 (2013).

\bibitem {k}A. Kempf, J. Phys. A. \textbf{30}, 2093 (1997).

\bibitem {3}L. N. Chang, D. Minic, N. Okamura, and T. Takeuchi, Phys. Rev. D
65, 125027(2002).

\bibitem {garay}L. J. Garay, Int. J. Mod. Phys. A\textbf{\ 10}, 145 (1995).

\bibitem {amati}D. Amati, M. Ciafaloni, and G. Veneziano, Phys. Lett.
B\textbf{\ 216}, 41 (1989).

\bibitem {m}M. Magiore, Phys. Lett. B 319, 83 (1993).

\bibitem {4}D. Bouaziz and M. Bawin. Phys. Rev. A 76, 032112 (2007).

\bibitem {5}W. Greiner, \textit{Relativistic Quantum Mechanics}, (3rd Edition
Springer-Verlag Berlin Heidelberg Germany 2000), p. 53.

\bibitem {bound state}F. D. Adame, Can. J. Phys. 67, 992 (1989).

\bibitem {6}D. Bouaziz and N. Ferkous. Phys. Rev. A 82, 022105 (2010).

\bibitem {case}K. M. Case, Phys. Rev. 80, 797(1950).

\bibitem {alha}A. D. Alhaidari, Int. J. Mod. Phys. A, 25, 3703 (2010).

\bibitem {ronv}A. Ronveaux, \textit{Heun's Differential Equations}. (Oxford
University Press, Oxford, England, 1995).

\bibitem {hon}M. N. Hounkonnou and A. Ronveaux, Generalized Heun and
Lam\'{e}'s equations: factorization, arXiv : math-ph/ 0902.2991 (2009).

\bibitem {wang}Z. X. Wang, D. R. Guo, \textit{Special functions}, (World
Scientific Publishing, 1989) p. 66.
\end{thebibliography}
\end{document}